\documentstyle[11pt,oldlfont]{article}
\def\Re{\rlap{\rm I}\mkern3mu{\rm R}}
\def\a{\alpha}

\def\g{\gamma}

\def\d{\delta}

\def\f{\varphi}

\def\k{\kappa}

\def\Si{\Sigma}

\def\Fb{\mbox{\boldmath$F$}}
\def\Hb{{\cal H}}
\def\Fc{{\cal F}}
\def\Bb{{\cal B}}
\def\Mc{{\cal M}}
\def\pb{\mbox{\boldmath$\psi$}}

\def\gab{\mbox{\boldmath$\g$}}
\def\ffb{\mbox{\boldmath$f$}}

\def\st{\mbox{\boldmath$*$}}

\def\S{\mbox{\boldmath$\Sigma$}}

\def\X{\mbox{\boldmath$\Theta$}}
\def\Xb{\mbox{\boldmath$X$}}

\def\Ab{\mbox{\boldmath$A$}}
\def\Kb{\mbox{\boldmath$K$}}
\def\Yb{\mbox{\boldmath$Y$}}

\def\eg{\mbox{\boldmath$e$}}
\def\eb{\mbox{\boldmath$e$}}
\def\Sg{\mbox{\boldmath$S$}}

\def\Eb{\mbox{\boldmath$E$}}
\def\Hbb{\mbox{\boldmath$H$}}
\def\Bbb{\mbox{\boldmath$B$}}
\def\fb{\mbox{\boldmath$\f$}}
\def\ub{\mbox{\boldmath$\upsilon$}}

\def\th{\mbox{\boldmath$\theta$}}

\def\w{\mbox{\boldmath$\omega$}}
\def\we{\mbox{\footnotesize \boldmath$\wedge$}}
\def\L{\mbox{\boldmath$L$}}
\def\Lc{\mbox{\boldmath$\cal L$}}

\def\J{\mbox{\boldmath$J$}}
\def\U{\mbox{\boldmath$U$}}

\def\Wb{\mbox{\boldmath$W$}}

\def\Db{\mbox{\boldmath$\Delta$}}

\def\i{\mbox{\boldmath$i$}}

\def\ex{\mbox{\boldmath$d$}}

\def\bi{\bibitem}

\begin{document}

\thispagestyle{empty}

\begin{flushright}   hep-th/0205072 \\
                    LPT-ENS 01/28 \\
                     AEI-2001-062  \end{flushright}

\vspace*{0.5cm}

\begin{center}{\LARGE {On covariant phase space methods}}

\vskip1cm

Bernard Julia$^a$  and Sebasti{\'a}n Silva$^b$

\vskip0.5cm

$^a$Laboratoire de Physique Th{\'e}orique CNRS-ENS\\ 
24 rue Lhomond, F-75231 Paris Cedex 05, France\footnote{UMR 8549 du
CNRS et de l'{\'E}cole Normale Sup{\'e}rieure.
This work has been partly supported by the EU TMR contract
ERBFMRXCT96-0012 and HPRN-CT-2000-122. 
}\\
\vskip0.2cm
$^b$Max Planck Institut f{\"u}r Gravitationsphysik, Albert Einstein Institut,\\
Am M{\"u}hlenberg 5, D-14476 Golm, Germany

\vskip0.5cm

\begin{minipage}{12cm}\footnotesize

{\bf ABSTRACT}

\bigskip 

It is well known that the Lagrangian and the Hamiltonian formalisms can be
combined and lead to ``covariant symplectic''
methods. For that purpose  a ``pre-symplectic form'' has
been constructed from the
Lagrangian using the so-called Noether form. However, analogously to the
standard Noether currents, this symplectic form is only
determined up to total 
divergences which are however essential ingredients in gauge 
theories. 

We propose a new definition of the symplectic form which is covariant
and free of ambiguities in a general first order formulation. 
Indeed, our construction
depends on the equations of motion but not on the
Lagrangian. We then define a generalized Hamiltonian which generates
the equations of motions in a covariant way. 
Applications to Yang-Mills, general relativity,
Chern-Simons and supergravity theories are given. We also consider
nice sets of possible boundary conditions that imply the closure and 
conservation of the total symplectic form.

We finally revisit the construction of conserved charges associated
with gauge symmetries, from both the ``covariant symplectic''  and
the ``covariantized Regge-Teitelboim'' points of view. We find that
both constructions coincide when the ambiguity in the Noetherian
pre-symplectic form is fixed using our new prescription.
We also present a condition of integrability of the equations that lead
to these quantities.

\bigskip
\end{minipage}
\end{center}
\newpage

\section{Introduction}

Field theories are usually described by choosing between either a 
Lagrangian or an Hamiltonian formalism. In one case, Lorentz
covariance is manifest and the charges are computed via the Noether
procedure. On the other side, the Hamiltonian mechanics provides a
symplectic structure, with an associated bracket. However, this
framework requires a full 
machinery involving a space/time splitting, followed by a
Legendre transform and eventually by the identification of first and
second class constraints.

It is naturally tempting to look for a formalism which could combine
both, the explicit covariance and  symplectic geometry. A decisive
step in that direction was introduced in references
\cite{WCZ,ABR,LW,JK} where the
covariant symplectic method was developed. Using the so-called Noether
construction, a symplectic structure was computed in an explicitly
covariant way without the need for canonical momenta.

As long as the boundary contributions can be neglected, the
Noether method proposed in references \cite{WCZ,ABR,LW,JK} is
perfectly suitable and leads to a  well-defined symplectic form.
However, in most physically relevant field theories, boundary terms
are quite important. In these cases, the Noetherian approach fails to
produce a non-ambiguous symplectic structure without an additional
prescription. Indeed, the boundary contributions to the symplectic
structure cannot be computed in a canonical way, in complete  analogy
with the Noether symmetry currents which are only defined up to total
derivatives. We then recall in section \ref{prehal} the Noether
construction, we pay  special 
attention to the {\it
ambiguities}, we separate the {\it on-shell} from {\it off-shell}
statements 
and explain the role played by boundary terms (we treat the general case where
spacetime may have several boundary components).

In  section \ref{offshe} we propose a new method that is free of ambiguities
to construct the symplectic structure. The first step is to introduce auxiliary
fields such that that the equations of
motion of the theory become first order (no more than one
derivative). These auxiliary fields are the covariant generalization
of the canonical  
momenta of Hamiltonian mechanics. We then give an explicit formula for
the symplectic density which is covariant by construction and
depends only on the equations of motion. We compare our proposal with the
usual Noether 
construction in section \ref{compariso}. In the next subsection
\ref{hameq}, we follow the analogy with the ordinary Hamiltonian
formalism by defining a natural Hamiltonian
density which generates the dynamics. In other words, we give a
covariant generalization of $\partial_{t} q^{i}=\{H, q^{i}\}$.
In section \ref{eddd}, we compute the pre-symplectic
structure of 
Yang-Mills, gravity, (non-Abelian) Chern-Simons and eleven dimensional
supergravity theories. A careful comparison between our proposal and
the Noether construction of the symplectic density is carried out
on these examples. 
Finally, global considerations as the definition, the conservation and
the closure of the symplectic
form are studied in section \ref{bdcocl}. The relations with the
boundary conditions and the variational principle are also explained.

The construction of conserved charges associated with gauge symmetries
using the covariant symplectic methods is discussed in section
\ref{gsycc}. In section \ref{supsy}, we recall the standard method
used for instance in references
\cite{IW,AAA,Carlip:1999cy,WZ,CN,Francaviglia:2001ww}. The alternative
proposal of
reference \cite{Si1} is explained in section \ref{diffcri}. We then
show in section \ref{combre} that both constructions are equivalent
provided our new symplectic density is used. Finally, we comment on the
integrability of the ``covariantized'' Regge-Teitelboim equation
in section \ref{cobound}, this was a completely open question at
least until some result in \cite{WZ}.

In the body of this manuscript, we use a compact
differential form
notation. However, the important formulas and many
(technical) proofs can be found in the
appendix in  components.

\section{Some new results in the covariant
phase space formalism}\label{geopre}

We denote by $\Fc$ the space of smooth field configurations (off-shell) which satisfy
some given boundary conditions. The
exterior derivative on this infinite dimensional manifold is called
$\delta$. The spacetime manifold of dimension $D$ and its exterior
derivative are 
respectively denoted by $\Mc$ and $\ex$ (for more details on the
geometry of $\Mc$, see section \ref{bdcocl}). We take the usual convention
that the two exterior derivatives commute. 

In order to remove some ambiguities in the notation, one may distinguish
field configuration Forms 
with a capital F and spacetime forms with a lower case one. Moreover, all
spacetime differential forms are denoted with {\bf bold} characters.
A (p,q)-Form will be 
a p-form over $\Mc$ and a q-Form on $\Fc$. Finally, the wedge-product
$\we$ antisymmetrizes both, the spacetime indices ($\mu ,
\nu , \rho , \dots $) and the field configurations ones ($i, j, k, \dots $).

\subsection{The (off-shell) Noetherian  pre-symplectic forms}\label{prehal} 

We denote by $\fb^{i}$ the fields of a given theory. Generically,
the index $^{i}$ will be a Lie-valued index or will label different types
of fields, even auxiliary ones. For instance, in the first order
vielbein formulation of gravity, the index $^{i}$ encompasses the Lorentz
indices $(\  ^{a} ,\  ^{a}{}_{b})$ associated respectively with the
vielbein and the spin connection. The spacetime indices are taken into 
account by the use of differential forms, the degree of $\fb^{i}$ is
denoted by $p_{i}$. 

The dynamics is encoded in a $D$-form Lagrangian $\L$, which is a local
functional of the fields $\fb^{i}$. Under an arbitrary variation
$\fb^{i} \rightarrow \fb^{i} + \delta \fb^{i}$, the 
Lagrangian transforms into the equations of motion plus a
boundary term:
\begin{equation}\label{lagelv}
\delta \L = \delta \fb^{i} \we  \Eb_{i} + \ex \th.
\end{equation}
The Euler-Lagrange  variations (alias the variational derivatives of
the action, that is the equations of motion)
associated with the fields $\fb^{i}$ are  the
$(D-p_{i})$ spacetime forms $\Eb_{i}:=\frac{\delta \L }{\delta \fb^{i}}$.
The (D-1,1) Form $\th$ is called a { \it Noether-Form}, it is a 
$1$-Form in the space of field configurations $\Fc$ and a $(D-1)$-form
over spacetime $\Mc$.

Let us first assume that our Lagrangian $\L$ depends on the 
fields $\fb^{i}$ and at most on their first derivatives $\ex \fb
^{i}$. Then, the equations of motion are explicitly given by:

\begin{equation}\label{feqmot}
\Eb_{i} = \frac{\partial \L }{\partial \fb^{i}} -(-)^{p_{i}} \ex
\frac{\partial \L }{\partial \ex \fb^{i}}.
\end{equation}

This allows us to give 
a closed formula for a particular 
{\it Noether-Form} $\hat{\th}$ associated to  a given Lagrangian $\L$:
\begin{equation}\label{usufo}
\hat{\th} := \delta \fb^{i} \we \frac{\partial \L}{\partial \ex \fb ^{i}}.
\end{equation}

However, the general Noether-Form $\th$ is afflicted by two 
ambiguities \cite{LW,IW}:

\begin{enumerate}
\item [$\bullet$] From equation (\ref{lagelv}), we can add a
total derivative\footnote{We shall assume trivial cohomology in this work: 
otherwise the arbitrariness of $\th$
includes an arbitrary $\ex$-closed Form.} to $\hat {\th}$:
\begin{equation}\label{firstamb}
\th = \hat {\th} + \ex \Yb,
\end{equation}
for any $(D-2,1)$-Form $\Yb$.
\item [$\bullet$] The Lagrangian $\L'=\L + \ex \Kb$ generates
the same equations of motion as $\L$, that is, the bulk term in the
decomposition (\ref{lagelv}) remains unchanged. The associated
Noether-Form $\hat{\th}'$ 
(derived from equation (\ref{usufo})) however differs from $\hat{\th}$ by
$\delta$-exact and $\ex$-exact terms,
\begin{equation}\label{ambth}
\hat{\th}'  = \hat{\th} + \delta \Kb  + \ex \Yb (\Kb), 
\end{equation}
where we emphasized that in general the total derivative $\ex \Yb (\Kb)$
depends on the choice of $\Kb$ see for instance equation (2.15) of [3] .
\end{enumerate}

The relevant object is not the Noether-Form, but its exterior
derivative in ${\cal F}$, the so-called 
 {\it (off-shell) pre-symplectic (D-1,2)-Form}
\begin{equation}\label{prres}
\w_{\mbox{\it \tiny No}} := \delta \th.
\end{equation}
By definition (\ref{prres}), the pre-symplectic form  $\w_{\mbox{\it
\tiny No}}$ is $\delta$-closed. From the above ambiguities 
(\ref{firstamb}) and (\ref{ambth}), it is itself determined only up to a $\ex 
\delta$-exact term:
\begin{equation}\label{wrty}
\w_{\mbox{\it \tiny No}} = \hat{\w}_{\mbox{\it \tiny No}} + \ex \delta \Yb,
\end{equation}
where (see equation (\ref{usufo}))
\begin{equation}\label{hano}
\hat{\w}_{\mbox{\it \tiny No}}:= \delta \hat {\th} = - \delta \fb^{i} \we\delta \left( \frac{\partial
\L}{\partial \ex \fb ^{i}} \right).
\end{equation}

When boundary terms are important, the above
ambiguity in $\Yb $   has to be resolved in some way. For instance,
$\Yb$ was fixed in lemma 3.1 of \cite{IW} using a ``gauge 
covariance'' criterion. 
We shall propose a more general definition of the
pre-symplectic Form (henceforth called symplectic density) in the following 
subsection.
Basically we shall relax the $\delta$-closure condition and replace it by 
$\delta$-closure up to a total divergence. This will allow us to solve the
ambiguity problem in a general way.

\subsection{The covariant (off-shell) symplectic two-Form densities in
first order theories}  \label{offshe}

Equation (\ref{lagelv}) is nothing but the $\delta$-exterior
derivative in $\Fc $ (see previous definitions) of the Lagrangian $\L$. 
Let us now take a (second) $\delta$-exterior derivative of equation
(\ref{lagelv})\footnote{Recall that $\delta^{2}=0$, $\delta (A\we
B)=\delta A \we B+
(-)^{p_{A}} A \we \delta B$, with $A$ a $p_{A}$-Form on $\Fc$.}:
\begin{equation}\label{var}
\delta \fb^{i}\we  \delta \Eb_{i}= \ex \delta \th =: \ex \w.
\end{equation}

We shall define a {\it symplectic density} $\w$ by
requiring $\w$ to be {\it any 
(D-1,2)-Form which
satisfies equation (\ref{var})}. This defines $\w$  
up to a $\ex$-exact Form $\ex \Xb$:
\begin{equation}\label{amfor}
\w = \hat{\w} + \ex \Xb.
\end{equation}
Note that a pre-symplectic Noether Form $\w_{\mbox{\it \tiny No}}$ is
nothing but a symplectic density which is $\delta$-closed.

Let us now assume that we are working with a {\it first order theory},
this being defined by the property that the 
equations of motion $\Eb_{i}$  depend on $\fb^{i}$ and/or on $\ex \fb ^{i}$
but not on higher derivatives of the fields. 
This is not really a limitation: it is in fact
always possible to add auxiliary fields to the Lagrangian to reduce
second order  equations of motion to
first order ones while preserving the gauge symmetries \cite{He,JS2}. 

The point in going to these first order formulations is that we can
find an explicit formula for one symplectic density. 
In fact,
in appendix A, equation (\ref{nsymmm}), we exhibit the special representative: 
\begin{equation}\label{newpre}
\hat{\w}:= \delta \fb^{i}\we \delta \fb^{j} \we \frac{1}{2} \hat{\w}_{ij},
\end{equation}
with
\begin{equation}\label{dnewp}
\hat{\w}_{ij}:= (-)^{p_{i}} \frac{\partial  \Eb_{i}}{\partial \ex  \fb^{j}}.
\end{equation}

The Form $\hat{\w}$ gives a ``bulk'' contribution common to {\it all}
symplectic densities, more correctly any symplectic density differs
from it by a total divergence $\ex \Xb$. In
particular, if we assume that $\Xb$ vanishes at the
boundaries\footnote{For instance in reference
\cite{LW}, boundary terms are set to zero assuming either appropriate
boundary conditions or that the fields are of compact support.}, the
choice of $\Xb $ will be irrelevant and equation
(\ref{dnewp}) gives an alternative
definition for the symplectic density which may or may not be equal 
to $\hat{\w}_{\mbox{\it \tiny No}}$.

The  symplectic density $\hat{\w}$ depends only on the
equations of motion and 
satisfies some important local properties:
\begin{enumerate}
\item [] (i) $\ex \hat{\w}=\delta\fb ^{i} \we \delta \Eb _{i}$;
\item [] (ii) it is antisymmetric in $i$ and $j$, that is
\begin{equation}\label{antisyd}
\hat{\w}_{ij}  = -(-)^{p_i p_j} \hat{\w}_{ji},
\end{equation}
\item [] (iii) $\hat{\w}$ is generally covariant.
\end{enumerate}

Property (i) is a direct consequence of equations (\ref{var}) and
(\ref{amfor}) but it can
also be proven from definition (\ref{dnewp}). The
antisymmetry property 
(\ref{antisyd}) also follows after a simple calculation given in
equation (\ref{demma}) in component notation.
There the first order formulation is crucial.
The third point (iii) 
follows from definition (\ref{dnewp}). In
fact, the $\ex$-derivative can be replaced by a covariant
derivative in first order theories. The covariance of
$\hat{\w}$ then follows from the covariance of the equations of
motion.

As opposed to the  Noetherian pre-symplectic Forms
$\w_{\mbox{\it \tiny No}}$ given by equation (\ref{prres}), our more general
$\w$'s indeed are not automatically $\delta$-closed.
In fact, a direct calculation shows that (see the result (\ref{toder}))
\begin{equation}\label{extdehat}
\delta \w= \ex \left( \frac{1}{3} \delta \fb ^{i} \we \frac{\partial
\hat{\w}}{\partial \ex \fb ^{i}} + \delta \Xb \right).
\end{equation}

We have thus extended the consideration of $\delta$-closed Noetherian
pre-symplectic Forms  to the more general definition of  symplectic
densities, which are  $\delta$-closed only up to $\ex$-exact terms.
Consequently, we increased the ambiguity from an arbitrary
$(D-2,1)$-Form $\Yb$ (equation (\ref{wrty})) to an arbitrary
$(D-2,2)$-Form $\Xb$  (equation (\ref{amfor})). On the other hand, we
are now able to give a formula for the ``bulk'' symplectic
density as a functional of the equations of motion without direct
dependence on the Lagrangian. 
The later usually depends on the boundary conditions considered while
the Euler-Lagrange variation does not. 
After these observations it is natural to {\it choose} 
 $\Xb=0$, independently of the boundary conditions. That is, we 
should {\it define} the 
symplectic density  of a first order theory by equation (\ref{dnewp}). 
We will justify in the following this very
special way of fixing the ambiguity (\ref{amfor}).

\subsection{Comparison between  $\hat{\w}_{\mbox{\it \tiny No}}$ and
$\hat{\w}$} \label{compariso}

The two-Form $\hat{\w}_{\mbox{\it \tiny No}}$ is explicitly
$\delta$-closed  but 
depends on a choice of Lagrangian, see equation (\ref{hano}). On
the other hand, $\hat{\w}$
given by equation (\ref{dnewp}) is independent
of the  
Lagrangian (and explicitly covariant) but it is not in general
$\delta$-closed, see 
equation (\ref{extdehat}).
These two symplectic densities are however closely  related
to each other.
In fact a straightforward calculation given in (\ref{protot}) shows that:
\begin{eqnarray}
\hat{\w} &=&  \hat{\w}_{\mbox{\it \tiny No}} + \frac{(-)^{p_{i}}}{2} \ex
\left(\delta 
\fb^{i}\we \delta \fb ^{j}\we \frac{\partial^{2} \L }{\partial \ex \fb
^{j} \we \partial \ex \fb ^{i}} \right) \nonumber\\
&=& \hat{\w}_{\mbox{\it \tiny No}} - \frac{1}{2}\ex \left(\delta \fb
^{i}\we \frac{\partial \hat{\th}}{\partial \ex \fb ^{i}} \right),\label{wghj}
\end{eqnarray}
where we used the definition (\ref{usufo}) in the second line.

Note that while each term on the right hand side of equation
(\ref{wghj}) depends in general
on the choice of the Lagrangian, the sum does not. The differences
between $\hat{\w}$ and $\hat{\w}_{\mbox{\it \tiny No}}$ will be analyzed
on specific examples in section \ref{eddd}. Note that even when they
coincide each formulation exhibits different properties.

Finally, equation (\ref{wghj}) also provides
an alternative formula for 
$\delta \hat{\w}$ (other than (\ref{extdehat})) which is explicitly
$\ex \delta$-exact: 
\begin{equation}
\delta \hat{\w} = \frac{1}{2} \ex \left(\delta
\fb^{i}\we \delta  \frac{\partial \hat{\th}}{\partial \ex \fb ^{i}}
\right)\label{sashkj} 
\end{equation}

\subsection{The covariant ``Hamiltonian'' equations}\label{hameq}

In the previous subsections, we derived two ``bulk'' symplectic
densities $\hat{\w}$ and $\hat{\w}_{\mbox{\it \tiny No}}$. 
Each of these two objects can be used to  rewrite the equations of motion in a
form familiar from Hamiltonian mechanics. Let us 
 define a ``Hamiltonian'' $D$-form by:
\begin{equation}\label{hamil}
\Hbb = \ex \fb^{i} \we  \frac{\partial \L}{\partial \ex \fb^{i}} - \L.
\end{equation}

This formula generalizes in an obvious way the well-known definition
of Hamilton. Note however that $\Hbb$ depends on the fields and on their first 
derivatives as well but that we do not Legendre transform to
conjugate momenta here as earlier
advocates of this approach like De Donder and Weyl did (see for instance
\cite{Kastrup:1982qq,Kanatchikov:2000jz} and references therein). 
A straightforward calculation gives the Euler-Lagrange
variation (also called variational derivative) of the Hamiltonian 
(\ref{hamil}) with respect to $\fb^{i}$:
\begin{equation}\label{heqmot}
\frac{\delta \Hbb }{\delta \fb^{i}} = (-)^{p_{i}} \ex \fb^{j} \we
\hat{\w}_{ij}  - \Eb_{i}.
\end{equation}
 
The proof is provided in the appendix, equation (\ref{elkk}), in
component notation. 
Note that it is precisely $\hat{\w}$ defined by
equation (\ref{dnewp}) which appears on the right
hand-side of the identity (\ref{heqmot}). That is, the choice  
$\ex \Xb = 0$ is automatically selected.
Moreover, the final equation (\ref{heqmot}) does not depend on the
ambiguity of adding a surface term to the Lagrangian or to the Hamiltonian.
It is independent of the necessary choice of boundary conditions.

Another related and direct calculation
shows\footnote{Note that $\delta \Hbb = \delta \fb^{i} \we
\frac{\delta \Hbb }{\delta \fb^{i}} + \ex (\delta \fb^{i} \we
\frac{\partial \Hbb }{\partial \ex \fb^{i}})$.} that (see the result
(\ref{simeqf})) 
\begin{equation}\label{autrhamil}
\delta \Hbb  + \i_{\tiny \ex \fb} \hat{\w}_{\mbox{\it \tiny No}} = - \delta \fb^{i} \we \Eb_{i}
\end{equation}
where 
 $\i_{\tiny \ex \fb}$ denotes the interior product along the vector
$\ex \fb ^{i}$  (tangent to 
$\Fc$). More explicitly we have, 
\begin{eqnarray}
\i_{\tiny \ex \fb} \left(  \frac{1}{2} \delta \fb ^{i}\we \delta \fb
^{j}\we \Ab_{ij} \right) &=&
\ex \fb^{i} \we \delta \fb ^{j}\we \Ab_{ij} \label{intep} \\
\i_{\tiny \ex \fb} \ex
\left( \frac{1}{2} \delta \fb ^{i}\we \delta \fb ^{j}\we \Bbb_{ij}
\right) &=& -  \ex \left( \ex \fb^{i} \we 
\delta \fb ^{j}\we \Bbb_{ij} \right)\label{intep2}
\end{eqnarray}
for any $(p,0)$-forms $\Ab_{ij}$ and $\Bbb_{ij}$ 
(the minus sign in equation (\ref{intep2}) comes from usual
``differential-form gymnastics'').

Now it is precisely the pre-symplectic form $\hat{\w}_{\mbox{\it \tiny
No}}$ (see equation 
(\ref{hano})) which appears in equation
(\ref{autrhamil}). Although both terms on the left hand-side of this
formula  depend on
the choice of Lagrangian surface term their sum is
independent on it and is given by the equations of motion.

\bigskip

{\it In summary}, the dynamics of a field theory can be rewritten
equivalently in the following ways:
\begin{eqnarray}
0\approx -\delta \fb^{i}\we \Eb _{i} 
&=&  \delta \fb ^{i} \we \frac{\delta \Hbb }{\delta \fb^{i}} +
\i_{\tiny \ex \fb } \hat{\w}  \\
&=&  \delta \Hbb + \i_{\tiny \ex \fb} \hat{\w}_{\mbox{\it \tiny No}}.
\end{eqnarray}

\subsection{Examples}\label{eddd}

\subsubsection*{\underline{The Yang-Mills and gravity cases}}

Let us take the simplest but rather general example of first
order Lagrangian, 
\begin{equation}\label{fiodlag}
\L_{1} = \ex \fb^{i}\we \ffb_{i} + \L_{0},
\end{equation}
 with $\ffb_{i}$ and $\L_{0}$ some $(D-p_{i}-1)$-form and
$D$-form respectively which depend on the fields $\fb^{j}$ but not on
their derivatives. This kind of Lagrangian is sufficiently
general to cover (4d super)gravities, Yang-Mills or p-form
theories.

Since $\frac{\partial^{2} \L_{1}}{\partial \ex \fb^{i}\we \partial \ex
\fb^{j}}=0$, the symplectic densities $\hat{\w}$ and
$\hat{\w}_{\mbox{\it \tiny No}}$ should coincide (see equation
(\ref{wghj})). For 
illustrative purposes, let us check this explicitly:
Following our definition (\ref{dnewp}), a simple
calculation (for the Lagrangian
(\ref{fiodlag})) yields  
\begin{equation}\label{dght0}
\hat{\w}_{ij}=(-)^{p_{i}}\frac{\partial  \Eb_{i}}{\partial \ex  \fb^{j}} =
(-)^{p_i p_j} \frac{\partial \ffb_j}{\partial \fb^{i}}- \frac{\partial
\ffb_i}{\partial \fb^{j}} ,
\end{equation}
which obviously satisfy (\ref{antisyd}).

It is easy to compare with the result given by the Noether-Form method
(equation (\ref{hano})):
\begin{equation}\label{prefi}
\hat{\w}_{\mbox{\it \tiny No}}= \delta \left(\delta \fb^{i}\we \frac{\partial \L_{1}}{\partial \ex
\fb ^{i}} \right) = -\delta \fb^{i} \we \delta \ffb_{i}  = -\delta
\fb^{i} \we \delta \fb^{j} \we \frac{\partial \ffb_{i}}{\partial
\fb^{j}} = \hat{\w},
\end{equation}
where we used equation (\ref{newpre}) together with equation
(\ref{dght0}) for the last equality. 

As we argued in section \ref{offshe}, the result
(\ref{prefi}) will be automatically gauge invariant.
Let us consider more explicit examples:
For Yang-Mills theories, formula (\ref{prefi}) gives the
well-known result \cite{WCZ,LW,JK}:
\begin{equation}\label{ymc}
\hat{\w}_{\mbox{\tiny \it YM}}=\mbox{\it Tr}\left( \delta \Ab \we \st \delta \Fb \right)
\end{equation}

For the affine-$GL (D;\Re)$ gravity (see definitions and notations in
\cite{JS1,JS3}) we find:
\begin{equation}\label{glc}
\hat{\w}_{\mbox{\tiny \it GL}}=\frac{1}{16\pi G} \delta
\w^{a}_{\ c} \we  \delta 
\left( \sqrt{\left|g \right|} g^{cb} \S_{ab} \right) 
\end{equation}
with $\w^{a}_{\ c}$ and $g^{ab}$ the $GL (D;\Re )$ connection and
metric respectively,
$\S_{ab}=\frac{1}{(D-2)!}\varepsilon_{abc_{3}\dots c_{D}} \th^{c_{3}}\we \dots
\we  \th^{c_{D}}$  and $\th^{a}$ is the canonical one-form. 

From formula (\ref{glc}), we can extract the pre-symplectic Form in
the Palatini formalism by fixing $\theta^{a}_{\mu} = \delta ^{a}_{\mu}$
(this breaks all the $GL (D;\Re)$ gauge symmetry).  
In components the result is \cite{WCZ,ABR,LW,JK,IW}:
\begin{equation}\label{pala}
\hat{\omega}^{\mu }_{\mbox{\tiny \it Pa}} = \frac{1}{8\pi G} 
\delta \Gamma ^{[\mu }_{\nu \rho } \we \delta
\left( \sqrt{\left|g \right|}  g^{\nu ] \rho}  \right)  
\end{equation}
with $\Gamma ^{\mu }_{\nu \rho }= \Gamma ^{\mu }_{(\nu \rho )}$ the
torsionless connection. Note that we recover ordinary second order
Einstein gravity by using the metricity condition $\nabla_{\mu }
g_{\rho \sigma } = 0$ which implies $\delta \Gamma ^{\mu }_{\nu \rho }
= \frac{1}{2}g^{\mu \sigma } (\nabla_{\nu } \delta g_{\rho \sigma } +
\nabla_{\rho } \delta g_{\nu \sigma } - \nabla_{\sigma } \delta
g_{\nu \rho})$.

The result (\ref{glc}) also contains the symplectic density of the
first order orthonormal frame (vielbein) formalism. Indeed, 
after implementing the gauge
choice $g^{ab}=\eta ^{ab}$ which breaks $GL (D;\Re)$ down to $SO
(D-1,1;\Re )$, equation (\ref{glc}) reduces to 
\begin{equation}\label{viel}
\hat{\omega}^{\mu }_{\mbox{\tiny  \it SO}} = \frac{1}{8\pi G } \delta \omega ^{ab}_{\nu} \we
\delta \left( \left| e \right| e^{[\mu}_{a}e^{\nu ]}_{b} \right),
\end{equation}
with $e^{\mu}_{a}$ the inverse of the orthonormal frame and $\omega
^{ab}_{\mu}=\omega ^{[ab]}_{\mu}$ the $SO (D-1,1;\Re )$ spin connection.

\subsubsection*{\underline{Chern-Simons theories}}

Let us start with the five dimensional Abelian 
Chern-Simons theory. It is the simplest first order theory whose
Lagrangian is not of the form (\ref{fiodlag}):
\begin{equation}\label{lagcs}
\L^{\mbox{\tiny \it CS}_{5}} := \frac{1}{3} \Ab\we \ex \Ab \we \ex \Ab.
\end{equation}

The equations of motion are then simply:
\begin{equation}\label{eqmocs2}
\Eb^{\mbox{\tiny \it CS}_{5}} := \frac{\delta \L^{\mbox{\tiny \it
CS}_{5}}} {\delta \Ab } = \ex \Ab  \we \ex  \Ab.
\end{equation}

Following equation (\ref{newpre}), the symplectic density computed from
equation (\ref{eqmocs2}) gives
\begin{equation}\label{cspred}
\hat{\w}^{\mbox{\tiny \it CS}_{5}} = -\delta \Ab \we \delta \Ab \we \ex \Ab 
\end{equation}
which is gauge invariant as promised.

If we follow the Noether-Form construction recalled in
section \ref{prehal}, we get:
\begin{equation}\label{wolg}
\hat{\w}^{\mbox{\tiny \it CS}_{5}}_{\mbox{\tiny \it No}} = \delta \left(\delta \Ab \we \frac{\partial \L^{\mbox{\tiny \it CS}_{5}}}{\partial
\ex \Ab} \right)= -\frac{2}{3}\left(\delta \Ab \we \delta \Ab \we \ex \Ab
+ \delta \Ab \we \Ab \we \ex \delta \Ab  \right). 
\end{equation}

Note that $\hat{\w}^{\mbox{\tiny \it CS}_{5}}_{\mbox{\tiny \it No}}$ is not 
anymore explicitly gauge invariant. A 
straightforward calculation shows that it differs from (\ref{cspred})
by an exact term,
\begin{equation}\label{difccs}
\hat{\w}^{\mbox{\tiny \it CS}_{5}} - \hat{\w}^{\mbox{\tiny \it
CS}_{5}}_{\mbox{\tiny \it No}}= - \frac{1}{3} \ex
\left(\delta \Ab \we \delta 
\Ab \we \Ab  \right),
\end{equation}
in agreement with the general formula (\ref{wghj}).

Therefore, the gauge invariant result (\ref{cspred}) cannot be found
using the Noether-Form 
procedure because there is no $\Yb$ which could  cancel the right-hand
side of equation (\ref{difccs}). On the other hand, the symplectic density 
(\ref{cspred}) is not $\delta$-closed if no appropriate boundary
conditions are implemented (see subsection \ref{bdcocl}).

The higher-dimensional ($D=2n+1$) non-Abelian Chern-Simons case is
very similar. The equations of motion
are generally given by:
\begin{equation}\label{equmvgene}
\Eb^{\mbox{\tiny \it CS}_{2n+1}} = \Fb ^{n}
\end{equation}
where $\Fb :=\ex \Ab +\Ab \we \Ab$ is the curvature of a Lie-valued
gauge field and $\Fb ^{n} := \Fb\we \dots \we \Fb$ {\it (n times)}.

From equation (\ref{newpre}), we find the covariant
symplectic density of the theory 
\begin{equation}\label{scsss}
\hat{\w}^{\mbox{\tiny \it CS}_{2n+1}} = -\frac{n}{2}  \mbox{\it Tr} \left( \delta \Ab \we\delta \Ab
\we \Fb^{n-1}\right).
\end{equation}

On the other hand, the pre-symplectic Form given by the ``Noether method'' 
is not explicitly covariant as in the previous case. It is
moreover far more complicated to compute it because of the involved
structure of the non-Abelian Chern-Simons Lagrangian (see for instance
\cite{Si1} for  
explicit formulas and references therein).

\subsubsection*{\underline{Eleven-dimensional supergravity}}

As a last example which nicely combines the structures of gravity and
higher dimensional Chern-Simons theories, we compute the symplectic
structure of eleven dimensional supergravity \cite{CJS}. Moreover, it gives a
simple example with fermionic fields.

The complete first order formulation of eleven dimensional
supergravity was derived in 
\cite{JS2}. Partial results for the spin connection or for the
four-form gauge field treated as independent fields can be found
in \cite{CFGPN} and \cite{NVN} respectively.

We follow the differential forms notation of eleven dimensional
supergravity extensively 
detailed in \cite{JS2}. The independent fields are the elfbein
$\eg^{a}$, the three form $\Ab$, the Majorana gravitino $\pb$ and the
two auxiliary fields, namely the spin connection $\w^a_{\ b}$ and the
four-form field strength $\Fb$. 
Using again equation (\ref{newpre}), we find the following
symplectic density:
\begin{eqnarray}
\hat{\w}^{11} &=& \frac{1}{4\k^2} \delta \w^{ab} \we \delta
\eg^{c}\we \S_{abc} -\frac{i}{2}
\delta \bar{\pb}\we \gab_{(8)} \we \delta  \pb - \delta \Ab \we
\delta \st \Fb
 \nonumber\\
& & - \frac{i}{4} \delta \eg ^{a}\we \delta \bar{\pb} \we \left( 2
\gab_{(7)a} -\gab_{(6)} \we \eb_a \right) \we \pb
- \frac{i\kappa}{4} \delta \Ab \we
\delta \left(\bar{\pb}\we\gab_{(5)} \we \pb
\right) \nonumber\\
& & + \k \delta \Ab \we \delta \Ab \we \ex \Ab-\frac{i}{8} \delta
\eg^{a} \we \delta \eg^b \we  \bar{\pb} \we \left(  \gab_{(6)} \eta_{ab} + \gab_{(5) a}
\we \eb_{b} \right) \we \pb
\label{sugsymp}
\end{eqnarray}

This symplectic density is explicitly covariant and differs from the
Noether one in the
Chern-Simons term (first term of the last line), in analogy with equation 
(\ref{difccs}). 

Note that there is a non
vanishing symplectic coefficient between pairs of three form components  
(proportional to $\delta
\Ab \we \delta \Ab $) as well as pairs of elfbein's  (proportional to
$\delta \eg^{a} \we \delta \eg^b$). The first one arises because of
the Chern-Simons term in the action and the second reflects  the special
torsion term present in the Lagrangian of first order higher
dimensional supergravities  (see
the third term 
of equation (6.1) of \cite{JS2}). It is interesting that this torsion
term is needed for any supergravity of dimension five and more because
of $\gamma$-matrix gymnastics. The Chern-Simons
term is required for supersymmetry, also in five dimensions and
more. This suggests some deep relation between torsion and
Chern-Simons effects.

\subsection{Boundary conditions, conservation and closure of the
symplectic Form}\label{bdcocl}

\subsubsection*{\underline{The geometrical data}}

We denote by $\Mc$ a $D$-dimensional piece of spacetime between two
(partial) Cauchy hypersurfaces $\Sigma_{t_{1}}$ and
$\Sigma_{t_{2}}$. We suppose that $\Mc$ admits a foliation and thus is
topologically $\Sigma_{t}\times \Re$.  

Let  $\Mc$ be also bounded by a set of $n$
$(D-1)$-dimensional 
time-like or null hypersurfaces, denoted by $\Hb_{r}$, $r=\{1,\dots
,n \}$. 
Depending on our specific examples, these $\Hb_{r}$ can
be for example pieces of future (or
past) null infinity after some compactification \cite{Pe}, of spatial
infinity \cite{AR} or even of the
horizon of a (locally) isolated black hole \cite{AAA}. 

We can define at each time, a set
of $n$ $(D-2)$-dimensional closed manifolds embedded in $\Mc$ by 
$\Bb_{r} = \Si_{t} \cap \Hb_{r}$, $r=\{1,\dots ,n\}$. These $\Bb_{r}$
are simply all the disconnected boundary  components of $\Si_{t}$, that is
$\partial \Sigma_{t}=\sum_{r=1}^{n}\Bb_{r} $. We suppressed
an explicit time dependence for notational simplicity. When needed, we
will however write $\Bb_{r1}$ or $\Bb_{r2}$
for $\Bb_{r}$ at time $t_{1}$ or $t_{2}$ respectively.

We denote by $s_{t}: \Sigma _{t} \rightarrow \Mc$, $h_{r}: \Hb_{r}
\rightarrow \Mc$ and $b_{r}: \Bb_{r} \rightarrow \Mc$ the natural
embeddings in $\Mc$ of the submanifolds previously described. We will
use the shorthand notation $\int_{\Sigma _{t}} \w$ and
$\left. \w\right|_{\Sigma _{t}}$ for respectively $\int_{\Sigma _{t}}
s_{t}{}^{*} \w$ and $s_{t}{}^{*} \w$ (for any $\w$), and so on for
$\Hb_{r}, h_{r}$ and $\Bb_{r}, b_{r}$.

\subsubsection*{\underline{The symplectic structure, off-shell and on-shell}}

Let us first
define the {\it off-shell symplectic Form} by\footnote{We abusively use
the word ``symplectic'' as it is common in the literature although 
the quantity (\ref{prret}) is usually degenerate due to gauge
symmetries \cite{LW}.} 
\begin{equation}\label{prret}
\Omega_{t} := \int_{\Si_{t}} \w.
\end{equation}

{\it A priori}, this $(0,2)$-Form $\Omega_{t}$ depends on the choice
of the
(partial) Cauchy hypersurface 
$\Si_{t}$ {\it and} on the ambiguity on $\w$ (namely, the choice of
$\ex \Xb$ in equation (\ref{amfor})).

Let us denote by $\bar{\Fc}$ (subspace of $\Fc$) the space of
smooth fields {\it
that satisfy the equations of motion}\footnote{We adopt
the ``bar notation'' of reference \cite{LW} for on-shell
quantities. We therefore denote by a bar the pulled-back quantities on
$\bar{\Fc}$.}, namely $\Eb_{i} =0$. Since
we have a natural embedding $\epsilon$ of $\bar{\Fc}$ in $\Fc$, the 
two-Form $\w$ 
pulls back to a two-Form on $\bar{\Fc}$, namely $\bar{\w}=\epsilon
^{*}\w$. Note
that now the exterior derivative of  $\bar{\Fc}$, denoted by
$\bar{\delta}$ defines the linearized equations of
motion by
 $\bar{\delta} \Eb_{i}= 0$ on $\bar{\Fc}$.

We can now pull back equation (\ref{var}) (which is defined
on  $\Fc$) to $\bar{\Fc}$ to verify that
\begin{equation}\label{exw}
\ex \bar{\w}=0.
\end{equation}

By integrating this equation on $\Mc$, we find  
\begin{equation}\label{wwww}
\bar{\Omega}_{t_{1}}-\bar{\Omega}_{t_{2}}= \sum_{r=1}^{n}\int_{\Hb_{r}} \bar{\w}
\end{equation}
with the definition (\ref{prret}) naturally pulled back on $\bar{\Fc}$.

Then, the symplectic two-Form $\Omega_{t}$ will be on-shell conserved
only if the right-hand side of equation (\ref{wwww}) vanishes. 
This condition must be satisfied in order to define
a time-independent symplectic structure.

A natural {\it but stronger} requirement for the possible
boundary conditions on $\Hb_{r}$ would be:
\begin{equation}\label{verif}
 \left. \w \right| _{\Hb_{r}} =0 \ \ \forall r.
\end{equation}

This constraint can give in principle some indication on how to fix
the ambiguity (\ref{amfor}). Concretely, given a set of
boundary conditions on $\Hb_{r}$, we may want to find a Form $\Xb$ such
that equation (\ref{verif}) is satisfied. We shall argue momentarily
that the choice $\Xb = 0$ (and then $\w=\hat{\w}$) is appropriate if
the boundary conditions on $\Hb_{r}$ are compatible with a variational
principle.

Condition (\ref{verif}) also implies the
(off-shell) closure of the symplectic form (\ref{prret}),
\begin{equation}\label{clos}
 \delta \Omega = 0.
\end{equation}

This can be verified by the following argument. First, equation
(\ref{verif}) together with 
(\ref{extdehat}) imply that:
\begin{equation}\label{ikkop}
0=\int_{\Hb_{r}} \delta \w = \int_{\Bb_{r1}} \left( \frac{1}{3} \delta
\fb ^{i} \we \frac{\partial 
\hat{\w}}{\partial \ex \fb ^{i}} + \delta \Xb \right) -
\int_{\Bb_{r2}} \left(\mbox{\it ``same''} \right),
\end{equation}
for all times $t_{1}$ and $t_{2}$ and for all $r$'s
independently. Moreover, since each piece of the right hand-side of equation
(\ref{ikkop}) depends on {\it arbitrary} $\delta \fb^{i}$'s, each of
these two term has to vanish 
separately\footnote{In fact, it is enough to suppose that we can find
one $\delta \fb^{i}$ which vanishes at some $t_{1}$ but not at
$t_{2}$. Note that $\delta \varphi^{i}$ should be compatible
with the boundary conditions but does not need to satisfy the linear
equations of motion. In other words, $\delta \varphi^{i}$ belongs to
$T^{*} \Fc$ but not necessarily to $T^{*} \bar{\Fc}$.}, that is:
\begin{equation}\label{sewrt}
\int_{\Bb_{r}} \left( \frac{1}{3} \delta
\fb ^{i} \we \frac{\partial 
\hat{\w}}{\partial \ex \fb ^{i}} + \delta \Xb \right) = 0,
\end{equation}
where we again suppressed the time dependence for simplicity.

Using this result together with equation (\ref{extdehat}) again,
we straightforwardly find that the symplectic
Form is $\delta$-closed:
\begin{equation}\label{dcloooo}
\delta \Omega = \sum_{r=1}^{n} \int_{\Bb_{r}} \left( \frac{1}{3} \delta
\fb ^{i} \we \frac{\partial 
\hat{\w}}{\partial \ex \fb ^{i}} + \delta \Xb \right) = 0.
\end{equation}

\bigskip 

{\it In summary}, we have shown that: 
\begin{eqnarray}
\left. \w  \right|_{\Hb_{r}}=0 \ \ \forall r &\Rightarrow &
\partial_{t} \bar{\Omega} =0 \nonumber\\
&\Rightarrow & \delta \Omega = 0.\label{eeke}
\end{eqnarray}

\subsubsection*{\underline{The boundary conditions and the variational
principle}}  

We shall now relate the condition $\left.\hat {\w
}  \right|_{\Hb_{r}} = 0 $ with the boundary conditions imposed  on
$\Hb_{r}$ by the choice of space  $\Fc$.
Let us consider the boundary component $\Hb_{r}$, between two times $t_{1}$
and $t_{2}$. The 
boundary conditions on $\Hb_{r}$ lead to a well-defined variational
principle for an appropriate Lagrangian on our space $\Fc$ provided 
\begin{equation}\label{qpc}
\exists\   \hat{\th} \ 
\mbox{\it in the class (\ref{ambth}) such that}\
\left. \hat{\th} \right| _{\footnotesize \Hb_{r}}  = 0. 
\end{equation}

Note that in order to satisfy the variational principle, that is 
\begin{equation}\label{vpenfd}
\delta \Sg = \delta \int_{\Mc } \L = 0 \Leftrightarrow
\Eb_{i}=0 \ \ \mbox{\it (without any surface term),}
\end{equation}
the condition (\ref{qpc}) must hold on {\it each} boundary
$\Hb_{r}$ ($r=1,\dots ,n$) of the
spacetime $\Mc$ as well as on the Cauchy hypersurfaces $\Sigma_{t_{1}}$ and
$\Sigma_{t_{2}}$. This is the usual way to fix the boundary conditions of
a variational problem.

A direct consequence of the statement (\ref{qpc}) is that  
if the variational principle is satisfied on the
boundary $\Hb_{r}$, there will exist a pre-symplectic Form whose pullback
vanishes on this $\Hb_{r}$:
\begin{equation}\label{anww}
\mbox{(\ref{qpc})}\ \Rightarrow \  \exists\ \hat{\w}_{\mbox{\tiny \it
No}}:=\delta  \hat{\th},    \ \mbox{\it such that}\
\left.  \hat{\w}_{\mbox{\tiny \it No}} \right|_{\footnotesize 
\Hb_{r}} = 0.
\end{equation}

It is then possible to analyze the pullback of the symplectic density
$\hat{\w}$ (\ref{dnewp}) on the boundary $\Hb_{r}$. From
the identity (\ref{wghj}), 
we find that:
\begin{equation}\label{jklll}
\left. \hat{\w} \right|_{\footnotesize \Hb_{r}}  =  -
\left. \frac{1}{2}\ex \left(\delta \fb 
^{i}\we \frac{\partial \hat{\th}}{\partial \ex \fb ^{i}}
\right) \right|_{\footnotesize \Hb_{r}}.
\end{equation}

It is hard to prove that the righthand side of this equation will
vanish for all boundary conditions compatible with a variational
principle. We can however give a quite general argument: let us 
assume that our set of boundary conditions are denoted by:
\begin{equation}\label{bdden}
\left. \Fb^{A} (\fb) \right|_{\footnotesize \Hb_{r}} = 0,
\end{equation}
for $A=1,\dots N$, with $N$ a given number of boundary conditions.

We assume that $\Fb^{A} (\varphi)$ is a functional of the fields but
not of their derivatives. This is quite natural in first order
theories where the dynamical fields and the momenta (auxiliary fields) are
independent. We basically suppose that the boundary conditions are
functions of the canonical variables $(q,p)$ but not of
their derivatives. 

Now, since $\hat{\th}$ vanishes on $\Hb_{r}$ by hypothesis, it should take the
general form:
\begin{equation}\label{geformth}
\hat{\th} = \Fb^{A} \we \X_{A}^{0}+ \ex \Fb^{A} \we \X_{A}^{1} +
\delta \Fb^{A} \we \X_{A}^{2} + \ex \delta \Fb^{A} \we \X_{A}^{3} + O
(\Fb ^{2}),
\end{equation}
for some given functionals $\X_{A}^{0}, \X_{A}^{1}, \X_{A}^{2}$ and
$\X_{A}^{3}$. 

We can then check that the right hand-side of
equation (\ref{jklll}) vanishes on $\Hb_{r}$ for the general
expression (\ref{geformth}) by making use of equation (\ref{bdden})
and its $\delta$-derivative.

It is therefore tempting to extend our analysis to boundary conditions
more general than (\ref{bdden}), and therefore to 
conjecture that:

\begin{equation}\label{conj}
\mbox{(\ref{qpc})}\ \Rightarrow \left. \hat{\w} \right|_{\footnotesize
\Hb_{r}}  = 0.
\end{equation}
However, we were not able to prove this statement in complete generality.

\section{Gauge symmetries and conserved charges}\label{gsycc}

Let us now suppose that the Lagrangian $\L$ is invariant under the following
gauge symmetry that preserves the boundary conditions off shell:  
\begin{equation}\label{symtran}
\d_\xi \fb^{i} = \ex \xi^\a \we \Db_\a^{i} + \xi^\a \we \tilde{\Db}_\a^{i},
\end{equation}
with $\fb^{i}$ some $p_{i}$-form field, $\Db_\a^{i}$ and  
$\tilde{\Db}_\a^{i}$ some field-dependent $(p_{i}-1)$ and $p_{i}$
space-time forms respectively and $\xi ^{\alpha } (x)$ the
infinitesimal parameter. 

The Noether current associated with the gauge symmetry (\ref{symtran})
is given by (see \cite{JS1,Si1} and references therein)
\begin{equation}\label{fpst}
\J_\xi = \ex \U_\xi + \Wb_{\xi},
\end{equation}
with
\begin{equation}\label{defww}
\Wb_{\xi}:=\xi^\a \Db_\a^{i} \we \Eb_{i}\ \ \mbox{and}\ \ \Eb_{i}:=\frac{\d \L}{\d \fb^{i}}.
\end{equation} 

Equation (\ref{fpst}) means that the Noether current $\J _{\xi}$
is  on-shell the  divergence of a ``superpotential'' $\U_{\xi}$, in
other words it is weakly  
equal to a topological current. However, this
superpotential cannot be computed without a
``case by case'' prescription using the usual Noether method. In fact
Noether's theorem allows us to compute $\ex \J _{\xi}$ but not $\J
_{\xi}$ itself (see also the recent work \cite{Barnich:2001jy} and
references therein).  

Given a boundary  $\Bb_{r}$ (a closed $(D-2)$-dimensional manifold,
see section \ref{bdcocl}) our goal is to compute $\U_{\xi}$ in
an unambiguous way. This would allow us to define an associated charge
by:
\begin{equation}\label{charge}
Q_\xi^{(r)}:=\int_{\footnotesize \Bb_{r}}
\U_{\xi}. 
\end{equation}
An important point is that we can construct charges on each
boundary component $ \Bb_{r}$, $r=\{1,\dots ,n \}$, completely independent from
each other.

The symplectic methods described in the previous section can be used to
derive the superpotential $\U_{\xi}$, in analogy with the
Hamiltonian result of Regge and Teitelboim \cite{RT,BH}. Let us first recall
the construction developed by Ashtekar, Wald and collaborators
and the alternative method proposed in reference \cite{Si1}.
In the third subsection, we clarify in which cases they are equivalent.
Finally in subsection \ref{cobound}, we shall  give a necessary condition
for the existence of the charge $Q_\xi^{(r)}$.

\subsection{The charges from the symplectic method}\label{supsy}

The off-shell symmetry $\delta_{\xi}\fb^{i}$ given by equation (\ref{symtran})
restricts naturally to a vector field on  $\bar{\Fc}$  since
$\delta_{\xi} \Eb_{i}$ vanishes on-shell. We can contract the
symplectic density (which is a two-Form on $\Fc$)  along this vector:
\begin{equation}\label{inprod}
\w_{\xi}:=\w 
(\delta_{\xi} \fb, \delta \fb)= \i_{\delta_{\xi}} \w,
\end{equation}
with
$\i_{\delta_{\xi}}$ the interior product (in $\Fc$) with respect to
the vector $\delta_{\xi}\fb^{i}$ (see also equations
(\ref{intep}) and (\ref{intep2})). 

The total charge associated with the symmetry (\ref{symtran}) obeys  
\cite{ABR,LW,JK}
\begin{equation}\label{noetde}
\delta Q_{\xi} =\i_{\delta_{\xi}} \Omega =  \int_{\Sigma_{t}} \w_{\xi}.
\end{equation}
The charge $ Q_{\xi}$ will be conserved (on-shell) only if the symplectic
Form is also time-independent.

For a gauge symmetry (\ref{symtran}),
$\w_{\xi}$
(defined by equations (\ref{amfor}) and (\ref{inprod})) is 
{\it on-shell}  
a total derivative (see the Appendix, equation (\ref{cghwq2}) for a
proof):  
\begin{equation}\label{protde}
\w_{\xi} \approx  \ex \left( \xi^{\alpha} \Db^{i}_{\alpha }\we
\delta \fb^{j}\we \hat{\w}_{ij} + \delta_{\xi} \fb^{i}\we
\delta \fb ^{j}\we \Xb_{ij}\right) =: \ex \ub_{\xi}.
\end{equation}

This implies that the total charge (\ref{noetde}) will be on-shell a boundary 
term. Equivalently, $ Q_{\xi}$ is given by the integral of the Noether
current on a Cauchy hypersurface:
\begin{equation}\label{ccha}
Q_{\xi } = \int_{\Sigma_{t}} \J_{\xi} \approx \int_{\partial
\Sigma_{t}} \U_{\xi}  
\end{equation}
by virtue of equation (\ref{fpst}). 
Then, using equations (\ref{noetde}) and
(\ref{protde}), we obtain an on-shell equation for the boundary charges,
\begin{equation}\label{cauch0}
\delta Q_{\xi}^{(r)} := \int_{\footnotesize \Bb_{r}} \delta \U_\xi \approx
\int_{\footnotesize \Bb_{r}} \ub_{\xi}\ \ \  \mbox{\it for all } \Bb_{r}\mbox{\it's}.  
\end{equation}

At each boundary $\Bb_{r}$,
the equation (\ref{cauch0}) gives an equation for the
variation of the superpotential. 
As in the Regge-Teitelboim procedure, the last step is to
``integrate''
the equations (\ref{cauch0}) using the boundary conditions imposed on
$\Bb_{r}$ in order to determine  $\U_{\xi}$. Moreover,
from equations  (\ref{protde}) and (\ref{cauch0}), we see that it is
extremely important to fix the tensor\footnote{\dots or $\Yb$ if
we restrict ourselves to $\delta$-closed 
pre-symplectic Forms, 
see section \ref{prehal}.} 
$\Xb$ in a correct way if we want to compute appropriate charges.

As a final comment, in principle the equation (\ref{cauch0})
has to be integrated (and therefore has to be integrable!) on {\it all}
the boundaries $\Bb_{r}$ to ensure consistency of equation
(\ref{noetde}). But we will see in the following subsections that one
can work in each boundary component separately.

\subsection{The superpotential from the ``differentiability'' of the
Noether current}\label{diffcri}

We now recall the construction of the charges given in \cite{Si1}
and compare it with the symplectic method described above.

Without giving a prescription, the superpotential is completely
ambiguous. In other words,  both
$\J_{\xi}$ and $\U_{\xi}$ of equation (\ref{fpst}) are unknown. This
equation then only gives the Noether current in term of the
superpotential, but no more. However, we can still take the
$\delta$-exterior derivative of equation (\ref{fpst}). We then obtain a
(D-1,1)-Form equation in $\Fc$ (off-shell):  

\begin{equation}\label{vasta}
\d \J_\xi=  \d  \fb^{i} \we \frac{\d \Wb_\xi}{\d \fb^{i}} +  \ex\left( \d\U_\xi + \delta \fb^{i} \we \frac{\partial
\Wb_\xi}{\partial \ex \fb^{i}} \right),
\end{equation}
where we assumed for the last identity that $\Wb_\xi$ depends at most on first
derivatives of the fields (first order theories).

Let us now choose {\it one} $\Hb_{r}$ and integrate equation (\ref{vasta}) on this hypersurface, between two times $t_{1}$ and $t_{2}$: 
\begin{equation}\label{intvasta}
\int_{\footnotesize \Hb_{r12}} \d  \J_\xi -  \int_{\footnotesize \Hb_{r12}} \d  \fb^{i} \we \frac{\d \Wb_\xi}{\d \fb^{i}}  = \int_{\footnotesize \Bb_{r1}} \left( \d\U_\xi + \delta \fb^{i} \we \frac{\partial
\Wb_\xi}{\partial \ex \fb^{i}} \right) - \int_{\footnotesize \Bb_{r2}} \left( \mbox{\it ``same''} \right).
\end{equation}

In analogy with the Hamiltonian formalism \cite{RT,BH}, the next idea is to
impose a  (local) ``differentiability condition'' for the Noether 
current $\J_{\xi}$ on the boundary $\Hb_{r}$, namely
\begin{equation}\label{ivasta0}
\int_{\footnotesize \Hb_{r12}} \d  \J_\xi =  \int_{\footnotesize \Hb_{r12}}\d  \fb^{i} \we \frac{\d \Wb_\xi}{\d \fb^{i}}.
\end{equation} 
Note that this equation is evaluated on a fixed boundary component
$\Hb_r$ and not on a Cauchy hypersurface $\Si_t$. Therefore, the
knowledge of the data on {\it one boundary} is enough for what follows.

We will justify equation (\ref{ivasta0}) in the next
subsection.  
For the time being let us assume that it is
satisfied. Then, the sum of the two terms on the  right hand-side of
equation (\ref{intvasta}) vanishes. However, since $\delta \fb^{i}$ is
{\it arbitrary}, each term has to vanish
separately (see the analogous argument after equation (\ref{ikkop})),
that is: 
\begin{equation}\label{defu2}
\delta Q_{\xi}^{(r)} = \int_{\footnotesize \Bb_{r}}  \d\U_\xi =-
\int_{\footnotesize \Bb_{r}} \delta \fb^{i} \we \frac{\partial
\Wb_\xi}{\partial \ex \fb^{i}}
\end{equation}
which is the main equation given in \cite{Si1}. Examples for which
this equation was used successfully can be found in
\cite{Si1,HJS,SiT,JS3,Si2,JuP,Silva:2002jq}. 

As opposed to equation (\ref{cauch0}), the right hand-side of equation
(\ref{defu2}) is non ambiguous since it does not depend on any
specific choice of $\Xb $ or $\Yb$.

\subsection{Comparison between methods \ref{supsy} and
\ref{diffcri}}\label{combre}

Using the
following on-shell identity,
\begin{equation}\label{ohjk}
(-)^{(p_{i}+1) (p_{j}+1)} \xi^{\alpha}\Db^{i}_{\alpha } \we \frac{\partial \Eb_{i}}{\partial
\ex \fb^{j}} \approx\frac{\partial
\left(\xi^{\alpha}\Db^{i}_{\alpha}\we \Eb _{i}  \right)}{\partial
\ex \fb^{j}} = \frac{\partial \Wb_{\xi}}{\partial \ex \fb_{j}}
\end{equation}
and using equations (\ref{dnewp}) and (\ref{protde}), we find that
\begin{equation}\label{llalla}
\ub_{\xi} \approx  -\delta \fb^{i}\we \frac{\partial
\Wb_{\xi}}{\partial \ex \fb^{i}} + \delta_{\xi} \fb^{i}\we
\delta \fb ^{j}\we \Xb_{ij},
\end{equation}
up to some $\ex$-exact terms which does not contribute to the charge since
$\Bb_{r}$ is a closed manifold.

Therefore, comparing equations  (\ref{cauch0}), (\ref{defu2}) and
(\ref{llalla}),  the ``differentiability'' condition (namely equation
(\ref{ivasta0})) is on-shell equivalent
to the criterion $\Xb=0$ in the choice of our symplectic density. In
other words, this 
criterion implies that the symplectic density of a first order theory
{\it is} $\hat{\w}$, defined by formula
(\ref{dnewp}).

Note finally that this result could also be recovered using the
(off-shell) identity\footnote{Equation (\ref{idenba}) was also used in
reference \cite{JS2} to compute the transformation laws of the auxiliary fields in first order supergravities.} \cite{Si1,SiT}
\begin{equation}\label{idenba}
\frac{\d \Wb_\xi}{\d \fb^{i}} = \frac{\partial}{\partial \ex 
  \fb^{i}} \left(\delta_{\xi} \fb^{j} \we \Eb_{j}\right).
\end{equation}
in  the integrand of equation (\ref{ivasta0})
\begin{equation}\label{kjlp}
\delta \J_{\xi} = \delta \fb^{i} \we \frac{\delta \Wb _{\xi}}{\d
\fb^{i}} \approx \delta _{\xi } \fb ^{i} \we \delta \fb^{j} \we (-)^{p_{i}}
\frac{\partial \Eb _{i}}{\partial \ex \fb ^{j}} = \hat{\w}_{\xi},
\end{equation}
where we used the definition (\ref{dnewp}) in the
last equality.

If we compare the result (\ref{kjlp}) with equations (\ref{noetde})
and (\ref{ccha}), we again
conclude that the 
differentiability condition singles out the non-ambiguous (and
covariant) symplectic density $\hat{\w}_{\xi}$ .

\bigskip 

Up to now, we have proven that equation (\ref{defu2}) is equivalent to
the differentiability condition (\ref{intvasta}), which on-shell, is
simply
\begin{equation}\label{genarg}
\delta \int_{\Hb_{r}} \J_{\xi } = \int_{\Hb_{r}} \hat{\w}_{\xi}.
\end{equation}
We then see that nothing is required on the other
boundary components $\Hb_{s}, s\neq r$. This is the mathematical
realization of the physical idea
that the ADM mass can be defined at spatial infinity no matter
what ``happens'' in the bulk or on the other internal boundaries as for
instance on a  horizon.

Now, equation (\ref{genarg}) can be  trivially justified if:
\begin{enumerate}
\item [$1$.] The charge $Q_{\xi}^{ (r)}=\int_{\Bb_{r}} \U _{\xi}$ is conserved.
\item [$2$.] The boundary conditions on  $\Hb _{r}$ are compatible
with the variational principle. 
\end{enumerate}

The point $1$ implies the vanishing of the left-hand side of equation
(\ref{genarg}) while the point $2$ implies the vanishing of the
right-hand side:
First, if the charge $Q_{\xi}^{(r)}$ is on-shell
conserved, then the flux of the Noether current vanishes on  $\Hb
_{r}$ due to equation  (\ref{fpst}):
\begin{equation}\label{peqrela}
\int_{ \Hb_{r}}\J_\xi = 0. 
\end{equation}
This implies in particular that 
$\int_{ \Hb_{r}}\delta \J_\xi = 0 $ and therefore the
left hand-side of equation (\ref{genarg}) is zero. 

Second, we argued in section \ref{bdcocl} that boundary conditions
adapted to a variational principle imply that $\left. \hat{\w}
 \right|_{\footnotesize \Hb_{r}} = 0$. This condition contracted along
the symmetry $\delta 
_{\xi} \varphi^{i}$ (a vector of $\Fc$) forces the vanishing of the
right-hand side of the equation (\ref{genarg}).

\bigskip 

{\it In summary}, we have first shown that equation (\ref{defu2})
is equivalent to the method of section \ref{supsy} if the
used Noetherian pre-symplectic form  is equal to the symplectic
density given by equation (\ref{dnewp}). Second, we have derived 
equation (\ref{defu2}) from the
natural assumptions 1 and 2 described above.

\subsection{Integrability}\label{cobound}

One can analyze the integrability of equations (\ref{noetde}) or (\ref{defu2})
 by taking their exterior $\delta$-derivative. This gives 
a necessary condition for the existence of the
charge $Q_{\xi }^{(r)}$.

As shown by equation (\ref{intres}) of the appendix, equation
(\ref{defu2}) will be integrable in $\bar{\cal F}$ if and only if
\begin{equation}\label{osinte}
\int_{\Bb_{r}} \left(\delta \fb^{k}\we \frac{\partial \delta _{\xi } \fb ^{
i}}{\partial \ex \fb ^{ k}} \we \delta \fb ^{j}\we
\hat{\w}_{ij}-\frac{(-)^{(p_{i}+1) (p_{j}+p_{k})}}{2} \delta _{\xi }
\fb ^{ k} \we \delta \fb ^{i}\we \delta \fb  ^{j} \we \frac{\partial
\hat {\w }_{ij}}{\partial \ex \fb ^{k}}  \right) \approx 0.
\end{equation}

The second term  of equation (\ref{osinte}) comes from the failure of
$\hat{\w}$ to be locally $\delta$-closed. However, this term will
vanish for suitable boundary conditions  when integrated on $\Bb
_{r}$. This follows after contracting the equation
(\ref{sewrt}) along a symmetry $\delta_{\xi }\fb ^{ i}$ (remember also
that $\Xb =0$).

Therefore, for boundary conditions compatible with the variational
principle, the conserved charge $Q^{ (r)}_{\xi }$ will exist only if
\begin{equation}\label{finintcon}
\int_{\Bb_{r}}\delta \fb^{k}\we \frac{\partial \delta _{\xi } \fb ^{
i}}{\partial \ex \fb ^{ k}} \we \delta \fb ^{j}\we
\hat{\w}_{ij} \approx 0.
\end{equation}

This condition is trivially satisfied for any Lie-type gauge
symmetry. In fact, in that case the symmetry transformation laws do
not depend on the derivatives of the fields.
The example of a diffeomorphism invariant theory is more
interesting. In that case,
$\delta _{\xi } \fb ^{i} = \ex \i_{\xi }\fb ^{i}+ \i_{\xi }\ex \fb
^{i}$, with $\i_{\xi}$ the interior product along the vector $\xi^{
\mu}$. Then it is easy to check\footnote{This can be proven using
components. If fact, $\delta_{\xi } \f^{ i} =
\xi ^{\rho }\partial_{\rho } \f^{ i}  + \mbox{\it ``more''}$, where
{\it ``more''} does not contain derivatives of the fields. We can then
plug this transformation law into equation (\ref{intres}), and
Hodge-dualize to get the result (\ref{diffini}).} that equation
(\ref{finintcon}) 
reduces to
\begin{equation}\label{diffini}
\int_{\Bb_{r}} \i_{\xi} \hat{\w} \approx 0.
\end{equation}

 This result was found in reference
\cite{WZ} using a direct
calculation in general relativity. We have just proven that this
results indeed remains valid for any diffeomorphism-invariant theory.

We can also comment on the integrability condition of the symplectic equation
(\ref{noetde}):

\begin{equation}\label{nouve}
\delta \i_{\delta _{\xi} } \Omega =\Lc_{\delta_{\xi}} \Omega =
\delta_{\xi} \Omega  = 0 
\end{equation}
where $\Lc_{\delta_{\xi}}=\delta \i_{\delta_{\xi}} +\i_{\delta_{\xi}}
\delta$ is the Lie derivative in $\Fc$ with respect to the vector
$\delta_{\xi}\fb^{i}$. 
In the second equality of equation (\ref{nouve}), we used the $\delta$-closure
of $\Omega$ (for appropriate boundary conditions, see equation
(\ref{eeke})). The third equality follows using that
$[\Lc_{\delta_{\xi}}, 
\delta]=[\Lc_{\delta_{\xi}},\ex]=[\delta_{\xi},\delta]=[\delta_{\xi},\ex]=0$.
We then see from equation (\ref{nouve}) that if the symplectic
structure  $\Omega$ is gauge invariant, the integrability is
guaranteed. 

Moreover, for a theory invariant under diffeomorphisms we can recover
the result 
(\ref{diffini}). In fact, if $\delta _{\xi}$
is
the Lie derivative
$\Lc_{\xi}$ along the vector field  $\xi^{\mu} (x)$, the equation
(\ref{nouve}) implies that $\int_{\partial \Sigma_{t}} \i_{\xi} \w
\approx 0$ since $\ex \w \approx 0$ by equation (\ref{exw}).

\section{Conclusion}\label{coooc}

We have proposed a new definition for the covariant symplectic form in
the Lagrangian framework. The associated symplectic density is covariant and
boundary condition independent since it is constructed using only the
equations of motion of the theory. For consistency we required
boundary conditions compatible with the variational principle. 
We have also shown that our
proposal coincides with the Noetherian pre-symplectic form of  
Yang-Mills and general relativity theories. However differences were
found for higher dimensional Chern-Simons theories and eleven
dimensional supergravity. 
We finally also defined a generalized Hamilton functional which together
with our symplectic density generates the dynamics.

In the second part of the paper, we have revisited the construction of
the conserved charges associated with gauge symmetries. We first recalled
the standard covariant symplectic method paying special attention
to ambiguities, on-shell and off-shell statements and to boundary
terms. We then concluded that this construction coincides with the one
proposed in \cite{Si1} if the ambiguity in the symplectic density
is fixed according to our new prescription.

\bigskip

{\bf Acknowledgments.}
\bigskip

We would like to thank C. Beetle for discussions
on Isolated Horizons.

\section*{Appendix: The symplectic density in components}\label{appa}
\renewcommand{\theequation}{A.\arabic{equation}}
\setcounter{equation}{0}

The purpose of this appendix is to give rigorous proofs of several
statements made in the main part of the manuscript. We work in
component notation, 
in the sense that spacetime forms are replaced by their duals, and the
spacetime indices, namely $\mu, \nu, \rho ,\dots $, are written
explicitely. Therefore, no bold characters (which denote in the main
text a spacetime differential form, see section \ref{geopre}) are used.
This gives a ``component translation'' of the main
results.

\subsection*{Basic definitions and notations}\label{basnot}

Let us start with some notations. We define the variational and
partial  derivatives:
\begin{equation}\label{pardef}
\delta_{i}:=\frac{\delta }{\delta \f^{i}} \ \
\partial_{i}:=\frac{\partial}{\partial \f ^{i}} \ \
\partial_{i}^{\mu}:=\frac{\partial}{\partial \partial _{\mu}\f ^{i}}.
\end{equation}

The operators (\ref{pardef}) act on functionals $A$ which 
depend on the fields $\f^{i}$ and their first derivatives only, that
is  $A=A
(\f,\partial_{\mu} \f )$. In that case, the Euler-Lagrange variation
can be rewritten as:
\begin{equation}\label{elv}
\delta_{i} A=\partial_{i} A-\partial_{\mu} \partial^{\mu}_{i}A.
\end{equation}

Using the notation (\ref{pardef}), the spacetime derivative operator
acting on a functional $A$ is simply:
\begin{equation}\label{dnoer}
\partial_{\mu} A = \partial_{\mu} \f^{i} \partial_{i} A +
\partial_{\mu} \partial_{\nu} \f^{i} \partial_{i}^{\nu} A.
\end{equation}

The variation of a functional $A$ (that is, its exterior derivative in
$\Fc$) under an arbitrary variation of the fields $\delta \f ^{i}$ can be rewritten in two natural ways:
\begin{eqnarray}
\delta A &=& \delta \f^{i} \partial_{i} A + \partial_{\mu}\delta
\f^{i}  \partial_{i}^{\mu} A \label{nvar1}\\
 &=& \delta \f^{i} \delta_{i} A + \partial_{\mu}\left(\delta \f^{i}
\partial_{i}^{\mu} A \right). \label{nvar2}
\end{eqnarray}

Note that $\delta$ (as $\partial_{i}$, $\partial_{i}^{\mu}$ and
$\partial_{\mu}$) is a derivation, in the sense that $\delta
(AB)=\delta (A)B+ A\delta (B)$ (with the usual sign correction if $A$
is an odd $p_{A}$-Form of $\Fc$). This is not true for $\delta_{i}$. A
straightforward calculation (using the definition (\ref{elv})) shows that
\begin{equation}\label{nonder}
\delta_{i}\left(AB \right) = \delta_{i}\left(A \right) B + A
\delta_{i}\left(B \right) - \partial_{\mu} A \partial_{i}^{\mu} B-
\partial_{i}^{\mu} A \partial_{\mu} B.
\end{equation}

A well-known result from variational calculus is that the
Euler-Lagrange variation of a derivative vanishes identically:
\begin{equation}\label{idvan}
\delta _{i} \partial_{\mu} A = 0 \ \ \ \forall A.
\end{equation}
Note that in this last equation, $A$ can carry any spacetime index.

In general, the basic operators $\partial_{i}$, $\partial_{i}^{\mu}$,
$\delta_{i}$  and  
$\partial_{\mu}$ (\ref{pardef}-\ref{dnoer}) do not commute (they however all
commute with $\delta$ (\ref{nvar1}-\ref{nvar2})). A straightforward
calculation using the above definitions shows that:
\begin{eqnarray}
\left[ \partial_{\mu},\partial_{i}^{\nu}\right] A &=& -\delta
^{\mu}_{\nu}\partial_{i} A \label{com1}\\
\left[ \partial_{\mu},\delta_{i}
\right] A &=& \partial_{\mu} \left(\delta _{i} A\right)  \label{com2} \\
\left[ \partial_{i}^{\mu},\delta_{j}\right]
A &=& -\partial_{i}\partial_{j}^{\mu} A  \label{com3}\\ 
\left[ \delta _{i},\delta_{j}
\right] A &=& \partial_{\mu}\partial^{\mu }_{j} (\delta_{i} A) +
\partial_{\mu} \partial_{\nu} (\partial_{i}^{\mu} \partial_{j}^{\nu}
A) \label{com4}\\
&=& \frac{1}{2} \partial_{\mu}\left( \partial^{\mu }_{j} (\delta_{i}
A)-\partial^{\mu }_{i} (\delta_{j} A)\right) \label{com4bis}\\
\mbox{\it and }\  \left[\cdot,\cdot \right] A &=& 0\ \ 
\mbox{\it for the others.} \label{com5}
\end{eqnarray}

Another useful formula for the following calculations is that 
\begin{equation}\label{uscal}
\delta_{j} \left(\delta _{i} A \right) = \delta
_{j}\left(\partial_{i}A \right) = \partial_{i} \left(\delta _{j} A \right)
\end{equation}
which comes from the definition (\ref{elv}), the property
(\ref{idvan}) and the commutation relations (\ref{com5}).

We will intensively use equations (\ref{pardef}-\ref{uscal}) in the following
subsections.

\subsection*{``Cascade equations'' for the symplectic density
2-Form; proof of equations (\ref{dnewp})}\label{casprof}

Let us start with
the (off-shell) equation (\ref{var}) which defines a symplectic density:
\begin{equation}\label{incovar}
\partial_{\mu} \omega^{\mu} = \delta \varphi^{i} \we \delta E_{i}.
\end{equation}

The symplectic density takes the general following
form (we work with first order theories): 
\begin{equation}\label{fffo}
\omega^{\mu}=\frac{1}{2}\omega^{\mu}_{ij} \delta \varphi^{i}\we \delta
\varphi^{j} + X^{\mu \nu }_{ij} \delta  \varphi^{i}\we
\partial_{\nu}\delta  \varphi^{j}.
\end{equation}

Therefore the equation (\ref{incovar})
together with (\ref{fffo}) and (\ref{nvar1}) can be rewritten as:
\begin{equation}\label{rewas}
\partial_{\mu }\left( \frac{1}{2}\omega^{\mu}_{ij} \delta \varphi^{i}\we \delta
\varphi^{j} + X^{\mu \nu }_{ij} \delta  \varphi^{i}
\we \partial_{\nu}\delta  \varphi^{j}\right) = \delta  \varphi^{i}\we
\left(\delta  \varphi^{j} \partial_{j} E_{i}+\partial_{\mu} \delta
\varphi^{j}  \partial_{j}^{\mu} E_{i}  \right).
\end{equation}

The equation (\ref{rewas}) contains a lot of information due to the
arbitrariness of $\delta  \varphi^{i}$ and of its derivatives. Then, analogously
to the ``cascade equations'' presented in the reference \cite{JS1}, the terms
proportional to $\delta \varphi^{i}\we \delta \varphi^{j}$, $\delta
\varphi^{i}\we \partial_{\mu } \delta \varphi^{j}$, etc\dots give
independent equations:
\begin{eqnarray}
\delta \varphi^{i}\we \partial_{\mu }\partial_{\nu } \delta
\varphi^{j} & &\left(  X_{ij}^{\mu 
\nu}  = 0  \right) \Rightarrow X_{ij}^{\mu\nu} =-  X_{ij}^{\nu\mu}
\label{cas1}\\ 
\partial_{\mu }\delta\varphi^{i}\we \partial_{\nu } \delta
\varphi^{j}  & & \left(  
X_{ij}^{\mu\nu}  = 0   \right) \Rightarrow X_{ij}^{\mu\nu} =
X_{ji}^{\nu\mu} \label{cas2}\\
\delta \varphi^{i}\we \partial_{\mu } \delta \varphi^{j} & & \left(\omega
^{\mu}_{ij}+\partial_{\nu }X^{\nu \mu }_{ij} = \partial_{j}^{\mu} E_{i} \right)\label{cas3}\\
\delta \varphi^{i}\we \delta \varphi^{j} & & \left(\frac{1}{2} \partial_{\mu }
\omega ^{\mu }_{ij} = \partial_{[j} E_{i]}=\delta _{[i}
E_{j]}\right). \label{cas4}
\end{eqnarray}
 
The first two equations simply show that $X^{\mu \nu }_{ij}=-X^{\nu
\mu }_{ij}=-X^{\mu \nu }_{ji}$. 
The third equation (\ref{cas3}), together with equation (\ref{fffo}),
shows that any 
symplectic density  which satisfies the equation (\ref{incovar})
can be rewritten as:
\begin{equation}\label{rewasssa}
\omega ^{\mu } = \hat{\omega}^{\mu } + \partial_{\nu} X^{\mu\nu },
\end{equation}
for some arbitrary $X^{\mu\nu }=\frac{1}{2}X^{\mu\nu}_{ij} \delta
\varphi^{i}\we\delta \varphi^{j}$ satisfying the above antisymmetry
properties. We also denoted by 

\begin{equation}\label{nsymmm}
\hat{\omega}^{\mu}= \frac{1}{2}\hat{\omega}^{\mu}_{ij} \ \delta\f ^{i}\we \delta\f
^{j} :=  \frac{1}{2} \partial_{j}^{\mu } E_{i}  \ \delta\f ^{i}\we \delta\f ^{j}, 
\end{equation}
the spacetime-Hodge-dual of equation (\ref{dnewp}).

Finally, it is straightforward to check that the last equation
(\ref{cas4}) follows after taking the divergence of (\ref{cas3}) (and
using equations (\ref{elv}) and (\ref{uscal})).

\subsection*{Antisymmetry of $\hat{\omega}^{\mu}_{ij}$}\label{antygg}

In this subsection we explicitly prove the antisymmetry property
(\ref{antisyd}) of the symplectic density $\hat{\omega}^{ \mu}$.

We assume that the equations of
motion of our theory are derived from a Lagrangian $L$, in
the class $L \sim L +\partial_{\mu} K^{\mu}$, by 
$E_{i}=\delta_{i} L$. The definition of
the symplectic density (\ref{nsymmm}) only depends on the equations of motion,
but not on a specific given Lagrangian. We then choose in the
above equivalence class one Lagrangian which depends at most on first derivatives of
the fields $L=L (\f, \partial_{\mu} \f)$ (for first order
theories). For that kind of 
Lagrangian, 
the equations
of motion are generally given by (see (\ref{elv}-\ref{dnoer})):
\begin{eqnarray}\label{equmot}
E_{i} = \delta_{i} L = \partial_{i} L -  \partial_{\mu}
\f^{j} \partial_{j}\partial^{\mu
}_{i} L  - 
\partial_{\mu} \partial_{\nu}
\f^{j} \partial_{j}^{\nu}\partial_{i}^{\mu} L.
\end{eqnarray}

Now, the restriction to {\it first order} theories, $E_{i}=E_{i} (\f,
\partial_{\mu }\f )$ implies that the tensor
\begin{equation}\label{tenan}
 \partial_{j}^{\nu}\partial_{i}^{\mu} L \ \
\mbox{\it is antisymmetric in $\mu$ and $\nu$},
\end{equation}
and then $E_{i} = \partial_{i} L -  \partial_{\mu}
\f^{j} \partial_{j}\partial^{\mu
}_{i} L$. Moreover, this condition implies 
that the tensor (\ref{tenan}) is also
antisymmetric in $i$ and $j$. 

We can then check the antisymmetry in $i$ and $j$ of the symplectic 
density (\ref{nsymmm}) by a straightforward calculation:
\begin{equation}\label{demma}
\partial_{j}^{\mu} E_{i} = \partial ^{\mu}_{j} \partial _{i}
L-\partial_{i}^{\mu}\partial_{j}L-\partial_{\nu}\f^{k}\partial_{k}
\partial_{j}^{\mu} \partial_{i}^{\nu}L = - \partial_{i}^{\mu} E_{j}   
\end{equation}
where the last equality follows from (\ref{tenan}).

\subsection*{Proof of equation (\ref{extdehat})}\label{prdhat}

Due to equations (\ref{incovar}) and (\ref{rewasssa}),
the symplectic density  $\hat{\omega}^{\mu}$ (\ref{nsymmm}) satisfies
the (off-shell) identity:
\begin{equation}\label{wrt}
\partial_{\mu } \delta \hat{\omega}^{\mu} = 0,
\end{equation}
that is from equation (\ref{nvar1}),
\begin{equation}\label{abb}
\frac{1}{2}\partial_{\mu}\left(\delta \varphi^{i}\we\delta
\varphi^{j}\we \left( \delta  \varphi^{k}
\partial_{k}\hat{\omega}^{\mu}_{ij} +   \delta \partial_{\nu} \varphi^{k}
\partial_{k}^{\nu}\hat{\omega}^{\mu}_{ij} \right) \right) = 0.
\end{equation}

We can then use the ``Abelian cascade trick'' given in the reference
\cite{JS1} to 
extract all the information contained in equation (\ref{abb}). This
basically consists 
in making the replacement $\delta \varphi \rightarrow \epsilon (x)
\delta \varphi$ in equation (\ref{abb})  (because of the 
arbitrariness of $\delta
\varphi$). The result is:
\begin{equation}\label{reggh}
\partial_{\mu } \left( \epsilon^{3} \delta \hat{\omega}^{\mu} +
\epsilon ^{2} \partial_{\nu} \epsilon \delta \varphi^{i}\we\delta
\varphi^{j}\we  \delta \varphi^{k} \partial_{k}^{\nu}\hat{\omega}^{\mu}_{ij}
\right) =0.
\end{equation}

Using the arbitrary of $\epsilon (x)$ and its derivatives, we find a
cascade of equations: 
\begin{eqnarray}
 \epsilon^{3} \partial_{\mu } \delta \hat{\omega}^{\mu}  &=& 0 \label{jca1}\\
\epsilon ^{2} \partial_{\mu } \epsilon \left( 3 \delta\hat{\omega}^{\mu} +
\partial_{\nu} \left( \delta \varphi^{i}\we\delta
\varphi^{j}\we  \delta \varphi^{k} \partial_{k}^{\mu}\hat{\omega}^{\nu}_{ij}
\right) \right)  &=& 0\label{jca2}\\
\left( 2 \epsilon \partial_{\mu }\epsilon  \partial_{\nu} \epsilon +
\epsilon ^{2} \partial_{\mu }\partial_{\nu }\epsilon \right) \left(
\varphi^{i}\we\delta  \varphi^{j}\we  \delta \varphi^{k}
\partial_{k}^{\nu}\hat{\omega}^{\mu}_{ij}\right) &=& 0\label{jca3}.
\end{eqnarray}

The first result (\ref{jca1}) obviously reproduces the equation
(\ref{wrt}) we
started with. The third equation (\ref{jca3}), together with
(\ref{nsymmm}) and (\ref{demma}), implies that
$\partial_{k}^{\nu}\hat{\omega}^{\mu}_{ij}=
\partial_{k}^{[\nu}\hat{\omega}^{\mu]}_{ij}=
\partial_{[k}^{\nu}\hat{\omega}^{\mu}_{ij]}$. 
Finally the second equations
simply says that the $\delta$-exterior derivative of
$\hat{\omega}^{\mu}$ is a total derivative:
\begin{equation}\label{toder}
\delta\hat{\omega}^{\mu} = \frac{1}{3}
\partial_{\nu} \left( \delta \varphi^{k} \we \partial_{k}^{\nu}
\hat{\omega}^{\mu} \right).
\end{equation}

Note that these results can also be obtained by a
direct (but tedious) calculation from the definition (\ref{nsymmm}).

\subsection*{Relation between
$\hat{\omega}^{\mu}_{\mbox{\it \tiny No}}$ and
$\hat{\omega}^{\mu}$}\label{relnot}  

We now prove equation (\ref{wghj}) by a straightforward calculation:
\begin{eqnarray}
\hat{\omega}^{\mu}_{\mbox{\it \tiny No}} &=& \delta \hat{\theta}^{\mu} \nonumber\\
&=& -
\delta \f^{ i} \we \delta \partial_{i}^{\mu} L \nonumber\\
&=& -\delta \f^{ i} \we \delta \f^{j} \partial_{j} \partial_{i}^{\mu}
L - \delta \f^{ i} \we \partial_{\nu} \delta \f^{j} \partial_{j}^{\nu}
\partial_{i}^{\mu} L \nonumber  \\
&=& \delta \f^{ i} \we \delta \f^{j}
\frac{1}{2}\left(\partial_{j}^{\mu} E_{i} -\partial_{\nu} \f^{k}
\partial_{k} \partial_{j}^{\nu}
\partial_{i}^{\mu} L\right)-\partial_{\nu}\left( \delta \f^{ i} \we
\delta \f^{j} \right) \frac{1}{2} \partial_{j}^{\nu}
\partial_{i}^{\mu} L \nonumber\\
&=& \hat{\omega}^{\mu } -\frac{1}{2}\delta \f^{ i} \we
\delta \f^{j} \partial_{\nu} \left( \partial_{j}^{\nu}
\partial_{i}^{\mu} L\right) +\frac{1}{2} \delta \f^{ i} \we
\delta \f^{j} \partial_{\nu }\partial_{\rho } \f^{k}
\partial_{k}^{\rho} \partial_{j}^{\nu} \partial_{i}^{\mu} L \nonumber\\
& &-\partial_{\nu}\left( \delta \f^{ i} \we
\delta \f^{j} \right) \frac{1}{2} \partial_{j}^{\nu}
\partial_{i}^{\mu} L \nonumber\\ 
&=&  \hat{\omega}^{\mu } -\frac{1}{2}\partial_{\nu }\left( \delta \f^{ i} \we
\delta \f^{j}  \partial_{j}^{\nu}
\partial_{i}^{\mu} L \right)\nonumber \\
&=& \hat{\omega}^{\mu } +\frac{1}{2}  \partial_{\nu}\left( \delta \f^{
i}\we \partial_{i}^{\nu} \hat{\theta}^{\mu} \right) \label{protot},
\end{eqnarray}
where we respectively used definitions (\ref{usufo}) and (\ref{hano}) and
equations (\ref{nvar1}), (\ref{demma}),
(\ref{nsymmm}), (\ref{dnoer}) and (\ref{tenan}).

\subsection*{The covariant ``Hamiltonian'' equations}\label{seha}

Let us now define the covariant Hamiltonian by:
\begin{equation}\label{hcov}
H:= \partial_{\mu} \varphi^{j} \partial_{j}^{\mu} L - L.
\end{equation}

The purpose is then to prove that the equations of motion $E_{i}$ are
equivalent to the ``covariant Hamilton equation'':
\begin{equation}\label{cha}
\delta_{i} H \approx \hat {\omega}^{\mu }_{ij} \partial_{\mu} \varphi^{j},
\end{equation}
where the symplectic density is given by our proposal (\ref{nsymmm})
and $\approx $ means on-shell.

To proceed, we can first check (using equations
(\ref{pardef}-\ref{uscal})) the following identity, valid on any
functional $A (\f,\partial_{\mu }\f)$:
\begin{equation}\label{commj}
\left[\delta _{i}, \partial_{\mu} \varphi^{j} \partial_{j}^{\mu}
\right] A = -2 \partial_{\mu }\partial_{\nu} \varphi^{j}
\partial_{i}^{\mu}\partial_{j}^{\nu} A.
\end{equation}

We then take the Euler-Lagrange variation $\delta _{i}$ of equation
(\ref{hcov}), which together with the equations (\ref{commj}),
(\ref{equmot}) and (\ref{tenan}), 
simplifies to:
\begin{equation}
\delta_{i} H 
= \partial_{\mu} \varphi^{j}\partial_{j}^{\mu} \left(\delta_{i}
L  \right) - \delta _{i} L. \label{elkk}
\end{equation}

Therefore, 
using the proposal (\ref{nsymmm}), the equation (\ref{elkk}) reduces to:
\begin{equation}\label{abohh}
\delta_{i} H = \hat{\omega}^{\mu}_{ij} \partial_{\mu} \varphi^{j} - E_{i}.
\end{equation}

This then proves that the Euler-Lagrange equations $E_{i}\approx 0$ can be
rewritten in an Hamiltonian form using our symplectic density 
(\ref{nsymmm}). Note finally that we can add an arbitrary total
derivative to the Hamiltonian (\ref{hcov}) (and of course to the
Lagrangian) without changing the 
equations of motion (\ref{cha}), again due to the identity
(\ref{idvan}).

\bigskip

As explained in section \ref{hameq}, the Noether pre-symplectic structure
$\hat{\omega}^{\mu}_{\mbox{\it \tiny No}}$ can also be used in order
to rewrite the equations of motion in an Hamiltonian form. Let us
contract equation (\ref{abohh}) with $\delta \f^{i}$ and then use
equation (\ref{nvar2}) on the left hand-side :
\begin{equation}\label{aumilyy}
\delta H -\partial_{\mu }\left(\delta \f ^{i} \partial_{i}^{\mu } H
\right) = -\i_{ \partial_{\mu}
\varphi} \hat{\omega}^{\mu} - \delta \f ^{i} E_{i}, 
\end{equation}
where the interior product (in $\Fc$) is defined as usual (see also
equations (\ref{intep}) and (\ref{intep2})):
\begin{eqnarray}
\i_{\partial_{\mu} \varphi} \left(\frac{1}{2} \delta \f^{i}\we \delta
\f^{j} A_{ij} \right) &=& \partial_{\mu} \varphi^{i} \delta
\f^{j} A_{ij}\label{intprodef}\\
\i_{\partial_{\mu} \varphi} \partial_{\nu} \left(\frac{1}{2} \delta
\f^{i}\we \delta 
\f^{j} B_{ij} \right) &=& \partial_{\nu} \left( \partial_{\mu}
\varphi^{i} \delta \f^{j} B_{ij}\right).\label{intprodef2}
\end{eqnarray}

Now, if we contract equation (\ref{protot}) along the vector (of $\Fc$)
$\partial_{\mu} \varphi^{i}$, we get that (see equation
(\ref{intprodef2})):
\begin{equation}\label{gyyui}
\i_{ \partial_{\mu}
\varphi} \hat{\omega}^{\mu} = \i_{ \partial_{\mu}
\varphi} \hat{\omega}^{\mu}_{\mbox{\it \tiny No}}+\partial_{\nu
}\left( \partial_{\mu} \f^{ i} \we
\delta \f^{j}  \partial_{j}^{\nu}
\partial_{i}^{\mu} L \right).
\end{equation}

Plugging this result into (\ref{aumilyy}) and using the explicit
definition of $H$ (\ref{hcov}) in the second term of the left hand, we
verify that this equation simplifies to:
\begin{equation}\label{simeqf}
\delta H = -\i_{ \partial_{\mu}
\varphi} \hat{\omega}^{\mu}_{\mbox{\it \tiny No}}- \delta \f ^{i} E_{i}.
\end{equation}

The equation (\ref{simeqf}) gives an alternative way to rewrite the
equations of motion. As opposed to equation (\ref{elkk}), this result
depends on the used Lagrangian, both in the Hamiltonian and in the
Noetherian pre-symplectic structure.

\subsection*{The symplectic density along a gauge symmetry}\label{djklll}

Let us assume that our theory is invariant under the following gauge
symmetry (see equation (\ref{symtran})):
\begin{equation}\label{gausmm}
\delta_{\xi } \varphi^{i} = \partial_{\mu } \xi ^{\alpha} \Delta ^{i
\mu}_{\alpha } + \xi ^{\alpha} \tilde{\Delta}^{i}_{\alpha }. 
\end{equation}

If we contract equation (\ref{incovar}) (which is satisfied by
$\hat{\omega}^{\mu}$) along the symmetry 
$\delta_{\xi } \varphi^{i}$, we get that on-shell:
\begin{equation}\label{odddd}
\partial_{\mu } \left( \hat{\omega}^{\mu}_{ij} \delta_{\xi }
\varphi^{i} \delta\varphi^{j} \right) \approx 0.
\end{equation}

We can again use the ``Abelian cascade equations'' technique
(equations (41-46) of reference \cite{JS1}) to extract the information
hidden in equation (\ref{odddd}), due 
to the arbitrariness of the gauge parameter $\xi ^{\alpha } (x)$. The
first step
is to replace $\xi ^{\alpha } (x)$ by $\epsilon (x) \xi ^{\alpha }
(x)$ in equation (\ref{odddd}):
\begin{equation}\label{fste}
\partial_{\mu } \left( \epsilon\  \hat{\omega}^{\mu}_{\xi } +
\partial_{\nu }\epsilon\  \Delta ^{i \nu }_{\alpha } \xi ^{\alpha}
\hat{\omega}^{\mu}_{ij} \delta\varphi^{j} \right) \approx 0, 
\end{equation}
where $\hat{\omega}^{\mu}_{\xi} := \i_{\delta_{\xi
} \varphi} \hat{\omega}^{\mu} = \hat{\omega}^{\mu}_{ij} \delta_{\xi }
\varphi^{i} \delta\varphi^{j}$ and we used equation (\ref{gausmm}).

Then, from the arbitrary of $\epsilon (x)$ and its derivatives, we
get a ``cascade'' of three equations:
\begin{eqnarray}
\epsilon & & \partial_{\mu }\hat{\omega}^{\mu}_{\xi }  \approx
0\label{cghwq1}\\ 
\partial_{\mu } \epsilon & & \left( \hat{\omega}^{\mu}_{\xi } +
\partial_{\nu } \left( \Delta ^{i \mu }_{\alpha } \xi ^{\alpha}
\hat{\omega}^{\nu}_{ij} \delta\varphi^{j}\right)\right) \approx
0\label{cghwq2}\\
\partial_{\mu } \partial_{\nu } \epsilon & & \left( \Delta ^{i \mu }_{\alpha } \xi ^{\alpha}
\hat{\omega}^{\nu}_{ij} \delta\varphi^{j}\right) \approx
0. \label{cghwq3}
\end{eqnarray}

As usual, the first equation (\ref{cghwq1}) reproduces the equation
(\ref{odddd}) we started with. The second equation (\ref{cghwq2})
shows that on-shell, $\hat{\omega}^{\mu}_{\xi }$ is a total
derivative. The last equation (\ref{cghwq3}) states that this total
derivative is antisymmetric in $\mu$ and $\nu$.

\subsection*{Integrability}\label{intoo}

As recalled in section \ref{diffcri}, the conserved charges associated with the
gauge symmetry (\ref{gausmm}) are computed on a boundary component
$\Bb_{r}$ by integrating the following equation:
\begin{equation}\label{bound}
\delta Q^{(r)}_{\xi} = -\int_{\Bb_{r}} \delta \varphi^{i}
\partial_{i}^{\nu} W^{\mu }_{\xi } d\Sigma _{\mu \nu }, 
\end{equation}
where
\begin{equation}\label{defwww}
W^{\mu }_{\xi } := \xi ^{\alpha } \Delta ^{i \mu }_{\alpha } E_{i}.
\end{equation}

Let us then analyze the integrability of equation
(\ref{bound}). Taking the 
 $\delta$-derivative, we find:
\begin{eqnarray}
\delta \left( \delta \varphi^{i}
\partial_{i}^{\nu} W^{\mu }_{\xi }\right) &=& - \delta \varphi^{i} \we
\partial_{i}^{\nu} \delta W^{\mu }_{\xi }\nonumber\\
&=& -\delta \varphi^{i} \we \partial_{i}^{\nu} \left(\delta
\varphi^{j} \delta_{j} W^{\mu }_{\xi } + \partial_{\rho } \left(\delta
\varphi^{j}\partial_{j}^{\rho }  W^{\mu }_{\xi } \right) \right),
\label{fiwrte} 
\end{eqnarray}
where we used equation (\ref{nvar2}) in the second line. 

Now the second term in the right hand-side of equation (\ref{fiwrte})
can be rewritten in the following way:
\begin{eqnarray}
 -\delta \varphi^{i} \we \partial_{i}^{\nu} \partial_{\rho } \left(\delta
\varphi^{j} \partial_{j}^{\rho }  W^{\mu }_{\xi } \right)  &=&
-\delta \varphi^{i} \we \left( \partial_{\rho } \partial_{i}^{\nu}\left(\delta
\varphi^{j}\partial_{j}^{\rho }  W^{\mu }_{\xi } \right) +
\partial_{i} \left( \delta
\varphi^{j}\partial_{j}^{\nu }  W^{\mu }_{\xi } \right)\right) \nonumber\\
&=& - \partial_{\rho } \left(\delta \varphi^{i} \we \delta \varphi^{j}
 \partial_{i}^{\nu} \partial_{j}^{\rho } W^{\mu }_{\xi } \right) +
\delta \varphi^{j} \we \partial_{\rho } \delta \varphi^{i}
\partial_{i}^{\rho } \partial_{j}^{\nu } W^{\mu }_{\xi }\nonumber\\
& & + \delta
\varphi^{j} \we \delta \varphi^{i} \partial_{i}\partial_{j}^{\nu }
W^{\mu }_{\xi } \nonumber\\
&=& \partial_{\rho } \left(\delta \varphi^{i} \we \delta \varphi^{j}
 \partial_{i}^{\rho} \partial_{j}^{\nu } W^{\mu }_{\xi } \right) +
\delta \varphi^{j} \we \delta \partial_{j}^{\nu }
W^{\mu }_{\xi }. \label{tobeplug}
\end{eqnarray}
where we used equations (\ref{com1}) and (\ref{nvar1}) and the
antisymmetry properties $\partial_{i}^{ \nu}
\partial_{j}^{ \rho} W_{\xi}^{ \mu } = \partial_{i}^{ [\nu}
\partial_{j}^{ \rho} W_{\xi}^{ \mu ] } = \partial_{[ i}^{ \nu}
\partial_{j]}^{ \rho} W_{\xi}^{ \mu }$ which follow from the identity 
$\partial_{i}^{ \nu} W_{\xi}^{ \mu } = \partial_{i}^{ [\nu}
W_{\xi}^{\mu ] }$ proven in the reference \cite{Si1}.

We can now plug the equation (\ref{tobeplug}) into equation
(\ref{fiwrte}) dropping the total derivative (first term
of equation (\ref{tobeplug})) since $\Bb _{r}$ is a closed manifold. 
The integrability condition then becomes:
\begin{equation}\label{inte}
\frac{1}{2} \int_{\Bb_{r}} \delta \varphi^{ i} \partial_{i}^{\nu }
\left(\delta \varphi^{j} \delta_{j} W_{\xi}^{ \mu } \right) d\Sigma
_{\mu \nu } = 0.
\end{equation}

We can now use the identity $\delta_{j} W_{\xi}^{ \mu } =
\partial_{j}^{\mu } \left( \delta _{\xi } \varphi^{ k} E_{k}\right)$
(see equation (\ref{idenba})) in the result (\ref{inte}), together
with the definition (\ref{nsymmm})  to verify
that the integrability condition on-shell becomes,
\begin{equation}\label{intres}
\int_{\Bb_{r}} \left( \delta \varphi^{i} \partial_{i}^{\nu }
\left(\delta _{\xi} \varphi^{ j} \right) \we \hat{\omega}^{ \mu}_{jk}
\delta \varphi^{k} - \frac{1}{2} \delta \varphi^{i} \we \delta
\varphi^{j} \delta _{\xi} \varphi^{k} \partial_{i}^{ \nu }
\hat{\omega}^{ \mu}_{jk}   \right) d \Sigma _{\mu \nu } \approx 0.
\end{equation}

The consequences of this result are analyzed in the
subsection \ref{cobound}.

\end{document}